\documentclass{aa}
\usepackage{amssymb,amsmath,natbib,xspace,graphicx,enumerate,ulem,centernot}
\usepackage[varg]{txfonts}
\bibpunct{(}{)}{;}{a}{}{,} 
\newcommand{\rau}{\ensuremath{(r/\mathrm{1\, AU})}}
\newcommand{\tstop}{t_\mathrm{s}}
\newcommand{\half}{\frac{1}{2}\xspace}
\newcommand{\OK}{\ensuremath{\Omega_\mathrm{K}}\xspace}
\newcommand{\ReN}{\ensuremath{\mathrm{Re}}\xspace}
\newcommand{\eg}{{e.g.\/}}
\newcommand{\ie}{{i.e.\/}}
\newcommand{\etal}{{et al.\/}}
\newcommand{\cf}{{c.f.\/}}
\newcommand{\desch}{\ensuremath{\text{Desch}}}
\newcommand{\mmsn}{\ensuremath{\text{MMSN}}}
\usepackage[%
pdfauthor={Elena Sellentin},
pdftitle={Sellentin \etal: Hydrodynamical effects on dust growth},
pdfstartview=FitH,
linkcolor=blue,
anchorcolor=blue,
citecolor=blue,
filecolor=blue,
menucolor=blue,
urlcolor=blue,
colorlinks=true]{hyperref}
\defcitealias{Sek}{ST03}
\begin{document}
\title{A quantification of hydrodynamical effects on \mbox{protoplanetary dust growth}}
\titlerunning{Hydrodynamical effects on dust growth}
\author{E.~Sellentin, J.~P.~Ramsey, F.~Windmark\thanks{Member of IMPRS for Astronomy \& Cosmic Physics at the University of Heidelberg}, C.~P.~Dullemond}
\authorrunning{Sellentin \etal}
\institute{Universit\"at Heidelberg, Zentrum f\"ur Astronomie, Institut f\"ur Theoretische Astrophysik, Albert-\"Uberle-Str.\ 2, D-69120 Heidelberg, Germany \ \ e-mail: \texttt{ramsey@uni-heidelberg.de}}
\date{\today}
%
%
\abstract
{The growth process of dust particles in protoplanetary disks can be modeled via numerical dust coagulation codes. In this approach, physical effects that dominate the dust growth process often must be implemented in a parameterized form. Due to a lack of these parameterizations, existing studies of dust coagulation have ignored the effects a hydrodynamical gas flow can have on grain growth, even though it is often argued that the flow could significantly contribute either positively or negatively to the growth process.}
{We intend to qualitatively describe the factors affecting small particle sweep-up under hydrodynamical effects, followed by a quantification of these effects on the growth of dust particles, such that they can be parameterized and implemented in a dust coagulation code.}
{Using a simple model for the flow, we numerically integrate the trajectories of small dust particles in disk gas around a proto-planetesimal, sampling a large parameter space in proto-planetesimal radii, headwind velocities, and dust stopping times.} 
{The gas flow deflects most particles away from the proto-planetesimal, such that its effective collisional cross section, and therefore the mass accretion rate, is reduced.  The gas flow however also reduces the impact velocity of small dust particles onto a proto-planetesimal. This can be beneficial for its growth, since large impact velocities are known to lead to erosion.  We also demonstrate why such a gas flow does not return collisional debris to the surface of a proto-planetesimal.}
{We predict that a laminar hydrodynamical flow around a proto-planetesimal will have a significant effect on its growth.  However, we cannot easily predict which result, the reduction of the impact velocity or the sweep-up cross section, will be more important.  Therefore, we provide parameterizations ready for implementation into a dust coagulation code.}
\maketitle
\label{introduction}
\noindent In the classical incremental growth scenario, planetesimals are formed by the coagulation of dust across several orders of magnitude in mass. During this growth phase, many different physical processes are important, but not all are beneficial. Indeed, several barriers have been found that stop dust particles from growing to sizes where gravity can aid in coagulation. For example, the charge barrier \citep{Okuzumi}, the radial drift and fragmentation barriers \citep{Naki, AccrWeid, Brauer, Birn}, or the bouncing barrier \citep{Zsom, WindB}.

In order to overcome these barriers, it is important to examine what effects neglected physical processes might have. One such effect is the hydrodynamical flow of disk gas around larger dust particles such as pebbles or small planetesimals, to which we will collectively refer to as proto-planetesimals. In this paper, we focus on the question of whether such a flow pattern is beneficial for the growth of protoplanetary dust. Although this question has already been partially addressed in \citet[hereafter \citetalias{Sek}]{Sek}, a quantification of these effects over a large parameter space such that the results can be implemented in a numerical dust growth code still lacks in the literature.

Whether colliding dust particles stick to each other and grow, bounce off each other, or disrupt one or both of the collision partners depends on a series of parameters, out of which the impact velocity is one of the most important. Experimental work has shown (for a summary, see \citealt{Makrowurm, Guettler}) that particles preferentially stick to each other if they collide at low velocities \citep{WurmStick}, and bounce off or disrupt at high collision velocities (\eg\ \citealt{Wurm25,Kotheetalastroph}). It was also found that larger dust aggregates are eroded due to high velocity impacts of dust monomers, a process that has become known as monomeric erosion \citep{Schraep}. Many of these experimental findings are included in the numerical simulations that model dust growth in protoplanetary disks (\eg\ \citealt{WeidA, TurbOrmel, Brauer, Birn, Zsom, Okuzumi2012}). Numerical studies also show that the growth of dust does not always proceed by a hierarchical coagulation of equal sized particles, but instead that collisions between particles of large size ratios can lead to the formation of planetesimals \citep{Xie, Wind}. In this scenario, proto-planetesimals can grow by sweeping up a secondary population of particles kept small by the collisional growth barriers.

To date, collision velocities in dust growth codes have been calculated from Brownian motion, radial and azimuthal drift, vertical settling, and turbulence induced motions of the dust particles. The collisional cross section of the particles is furthermore assumed to be equal to the geometrical cross section. This neglects potential effects of hydrodynamical gas flow past dust particles on the growth process. It is however expected that these effects can be quite important for collisions between particles of a large size ratio, \ie, in a sweep-up growth scenario.  Geometrical cross sections are only applicable if the trajectories of the colliding particles follow straight lines. This is not necessarily the case for a small dust particle passing near a large proto-planetesimal surrounded by a hydrodynamical flow pattern:  The drag of the gas flow around the proto-planetesimal can deflect the small dust particle, resulting in a curved trajectory. In the most extreme case, this deflection can prevent a collision, preventing the sweep-up of the dust particle. The collisional cross section must therefore be modified to take into account this deflection. 

The drag force of the gas flow on a small particle can also affect its speed. This then modifies the collision velocity between the dust particle and proto-planetesimal, which is decisive for determining whether a collision leads to sticking, bouncing, or erosion.

In protoplanetary disks, we expect hydrodynamical flow patterns to form around dust particles much larger than the mean free path of the gas because the gas orbits at sub-Keplerian speed while the dust orbits at Keplerian speed. This gives rise to a relative motion between disk gas and dust, appearing in the rest frame of the dust as a headwind. In this work, we aim to use numerical simulations to study the effect that such a flow pattern will have on the dust coagulation.

We assume that our proto-planetesimals are small enough so that gravity is negligible, and constraints on this assumption are presented below.  For studies which include the gravity of larger planetesimals, see, for \eg, \citet{Lambrechts2012,Morbidelli2012,Ormel2013}.

This paper is organized as follows: In Sect.~\ref{assumptions}, we present our assumptions and general properties of the flow pattern around a proto-planetesimal. In Sect.~\ref{effects}, we describe the changes to the collisional cross section and the impact velocity which we observe from integrating trajectories of small dust particles in the flow around a proto-planetesimal. In Sect.~\ref{reaccretion}, we turn to the question of whether gas flow around proto-planetesimals can return debris from a disruptive collision to the proto-planetesimal surface. In Sect.~\ref{conclusions}, we summarize our results.

\section{Validity range of the model}
\label{assumptions}
In the limit of high viscosity (\ie\ small Reynolds numbers), the steady-state ($\partial / \partial t = 0$) velocity field $\vec{v}_\mathrm{g}$ of a gas flow around a sphere of radius $R$ in spherical polar coordinates can be derived from the Navier-Stokes equations (\citealt{greinerstock}; originally due to Stokes):
\begin{equation}
\vec{v}_\mathrm{g}\, (r_\mathrm{s},\theta) = \vec{v}_\infty(\theta) \left(\frac{R^3}{4r^3_\mathrm{s}}\! +\! \frac{3R}{4r_\mathrm{s}}\! -\! 1\right) + \frac{3R}{4r^3_\mathrm{s}}(\vec{v}_\infty(\theta) \cdot \vec{r_\mathrm{s}}) \vec{r_\mathrm{s}}\left( 1\! -\! \frac{R^2}{r^2_\mathrm{s}}\right),
\label{eq:hydro}
\end{equation}
where $\vec{v}_\infty(\theta) = v_\infty \left(\cos \theta\, \hat{\vec{r}} - \sin \theta\, \hat{\vec{\boldsymbol{\theta}}}\right)$ is the upstream velocity, $v_\infty$ is the headwind velocity and constant, and $r_{\mathrm{s}}$ denotes the radial distance measured from the center of the sphere. As $r_\mathrm{s} \rightarrow \infty$, $\vec{v}_\mathrm{g}$ reduces to $-\vec{v}_\infty$. Eq.~(\ref{eq:hydro}) describes the flow around a spherical proto-planetesimal if a) the proto-planetesimal radius $R$ is much larger than the mean free path $\lambda$ in the disk, b) the flow is incompressible, c) laminar and unseparated, d) the gravity of the proto-planetesimal is negligible, and e) the proto-planetesimal does not rotate.

In the following, we assume (e) to be true and calculate when (a) -- (d) are valid for a protoplanetary disk in both the minimum mass solar nebula model (MMSN) \citep{WeidMMSN, Hayashi} and the model of \citet{Desch}.  Although the two models provide similar constraints, we quote the results for both.
\paragraph{Assumption a)}
Hydrodynamics is suitable to describe the macroscopic properties of a flow around a proto-planetesimal if $R \gtrsim 100 \lambda$. The mean free path $\lambda$ is related to the number density $n$ of gas molecules by $\lambda = 1/(n \sigma_\mathrm{g})$, where $\sigma_\mathrm{g}$ is the collisional cross section of the gas molecules. For $n$, we use the number density $n_0$ at the mid-plane of a protoplanetary disk, which is related to the surface density $\Sigma(r)$ by:
\begin{equation}
n_0(r) = \frac{\Sigma(r)}{\sqrt{2 \pi}\,h \mu m_\mathrm{p}},
\label{nmitte}
\end{equation}
with the disk scale height $h=c_\mathrm{s}/\OK$, and average molecular mass $\mu m_\mathrm{p}$, where $m_\mathrm{p}$ is the mass of a proton, $c_\mathrm{s} = \sqrt{k_\mathrm{B} T(r) / \mu m_\mathrm{p}}$ is the isothermal sound speed, $k_\mathrm{B}$ is the Boltzmann constant, and $\OK$ is the Keplerian angular frequency.

For the disk model, we assume $\mu = 2.3$ and $\sigma_\mathrm{g} = 2\cdot 10^{-19}~ \mathrm{m^2}$ \citepalias{Sek} for the collisional cross section of Hydrogen, and a stellar mass of $1\, M_\sun$. We adopt the disk temperature profile of \citet{Alexander}, which is consistent with observations:
\begin{equation}
T(r)= 100\, \rau ^{-1/2}~\mathrm{K},
\label{eq:temp}
\end{equation}
where $r$ is the heliocentric distance, expressed in astronomical units.  The gas surface densities for the \citet{Desch} and MMSN \citep{Hayashi} models are, respectively:
\begin{align}
\label{eq:surd}
\Sigma_{\desch}(r) &=  5\phantom{.} \cdot 10^5\phantom{0}\, \rau ^{-2.17}~\mathrm{kg\, m^{-2}}; \\
\label{eq:surm}
\Sigma_{\mmsn}(r) &= 1.7 \cdot 10^4\, \rau ^{-3/2}~\mathrm{kg\, m^{-2}}.
\end{align}
Substituting the surface densities into Eq.~(\ref{nmitte}), the minimal radius $R_\mathrm{min} = 100 \lambda$ that a proto-planetesimal must have for assumption (a) to hold is then:
\begin{align}
\label{eq:rminid}
R_{\min,\desch}(r) &= 3\cdot 10^{-2}\, \rau ^{3.42}\ \mathrm{m}; \\
\label{eq:rminim}
R_{\min,\mmsn}(r) &= 8\cdot 10^{-1}\, \rau ^{2.75}\  \mathrm{m},
\end{align}
which increases with heliocentric distance for both disk models.
\paragraph{Assumption b)}
Incompressibility approximately holds true for small Mach numbers, $M \lesssim 0.1$.
From the temperature profile (\ref{eq:temp}), the isothermal sound speed can be written as:
\begin{equation}
c_\mathrm{s}(r) = 600\, \rau ^{-1/4}\ \mathrm{m\, s^{-1}}.
\label{eq:csound}
\end{equation}
The headwind velocity $v_\infty = v_\mathrm{K}-v_\mathrm{g}$ originates from the sub-Keplerian rotation of the disk gas, and can be written as $v_\infty = \eta v_{\mathrm{K}}$ with \citepalias{Sek}:
\begin{equation}
\label{eq:eta}
\eta = -\frac{1}{2r\OK^2 \mu m_\mathrm{p} n_0}\frac{\partial P}{\partial r},
\end{equation}
where $P$ is the gas pressure as given by the ideal gas law.

Substituting the surface density profiles Eqs.~(\ref{eq:surd}) \& (\ref{eq:surm}), and the temperature profile Eq.~(\ref{eq:temp}) into the expression for $\eta$, we find the headwind velocity for the Desch and MMSN models to be
\begin{align}
\label{eq:vinfd}
v_{\infty,\desch} &\approx 24\, \mathrm{m\, s^{-1}}; \\
\label{eq:vinfm}
v_{\infty,\mmsn} &\approx 20\, \mathrm{m\, s^{-1}},
\end{align}
and independent of heliocentric distance.

From Eq.~(\ref{eq:csound}), the Mach number of a gas flow around a proto-planetesimal is thus on the order of $10^{-2}$ to $10^{-1}$ in both models for reasonable values of $r$, and assumption (b) is valid.
\paragraph{Assumption c)}
To determine whether the flow around a proto-planetesimal can be described as laminar, we first estimate the molecular viscosity of gas in a protoplanetary disk, which can be determined from:
\begin{equation}
\label{eq:viscosity}
\nu = \half\bar{u}_\mathrm{th}\lambda.
\end{equation}
Using Eq.~(\ref{eq:csound}), the mean thermal speed of the disk gas is given by $\bar{u}_{\mathrm{th}}(r) = \sqrt{8/\pi}\, c_\mathrm{s}(r)$. The mean free path can be calculated from Eq.~(\ref{nmitte}), and thus we have,
\begin{align}
\label{eq:viscid}
\nu_{\desch}(r) &= 0.13\, \rau ^{3.17}\, \mathrm{m^2s^{-1}};\\
\label{eq:viscim}
\nu_{\mmsn}(r)  &= 3.6\phantom{0}\, \rau ^{2.5}\phantom{^.}\, \mathrm{m^2s^{-1}}.
\end{align}
Based on experimental results, viscous hydrodynamic flow past a sphere will remain perfectly laminar and unseparated if the Reynolds number of the flow is $\ReN_\mathrm{crit}\! \lesssim\! 22$ (\eg\ \citealt{Taneda56}), where the Reynolds number of flow around a spherical proto-planetesimal is given by:
\begin{equation}
\label{eq:reynolds}
\ReN = \frac{2Rv_\infty}{\nu}.
\end{equation}
Other than the proto-planetesimal radius, all variables in Eq.~(\ref{eq:reynolds}) are now fixed by the choice of disk model. The statement that a flow will remain perfectly laminar below $\ReN_\mathrm{crit}\! \approx\! 22$ can then be used to solve for the maximal radius of a proto-planetesimal:
\begin{align}
\label{eq: RmaxD}
R_{\mathrm{crit},\desch} &= 0.059\, \rau^{3.17}\ \mathrm{m}; \\
\label{eq: RmaxM}
R_{\mathrm{crit},\mmsn} &= 1.98\phantom{..}\, \rau^{2.5}\ \mathrm{m}.
\end{align}
Laminarity therefore holds if the proto-planetesimal radius does not exceed $R_\mathrm{crit}(r)$.
\paragraph{Assumption d)}
We assume that the gravity of the proto-planetesimal is negligible if the drag force due to the friction of the fluid, $F_\mathrm{D}$, is at least $F_\mathrm{D} > 10^2 F_\mathrm{G}$, where $F_\mathrm{G}$ is the gravitational force of the proto-planetesimal.

The drag force can be written as $F_\mathrm{D} = mv/\tstop$ where $m, v$, and $\tstop$ are a dust particle's mass, speed, and stopping time. The largest stopping time that leads to a deflection in the flow pattern is approximately equal to the time $t_\mathrm{p}$ that a particle needs to be advected by the gas flow through the perturbed region around the proto-planetesimal. This timescale is thus $t_\mathrm{p} \approx L / v_\infty$, where $L$ is roughly the extent of the perturbed region. At a distance of $r_\mathrm{s}= 100\, R$, the flow pattern Eq.~(\ref{eq:hydro}) virtually equals the upstream velocity $v_\infty$. We therefore set the extent of the perturbed region to $L = 100\, R$ and thus $F_\textrm{D} = mv_\infty^2/100\, R$.

We evaluate the gravitational force at the surface of the proto-planetesimal by assuming early proto-planetesimals have a volume filling factor of $0.6$ \citep{GeretInhom}. We also assume that early proto-planetesimals consist of the same material as asteroids, which have typical densities of $3\cdot 10^3\, \mathrm{kg\, m^{-3}}$ \citep{Carry}. From this, we estimate the density of the proto-planetesimal to be $\rho_\mathrm{p} = 0.6 \times 3\cdot 10^3\, \mathrm{kg\, m^{-3}}$.

The statement $F_\mathrm{D} > 10^2 F_\mathrm{G}$ then becomes:
\begin{equation}
\label{eq:gravityineq}
\frac{mv_\infty^2}{100\, R} > 10^2\, \frac{G m \frac{4}{3}\pi R^3 \rho_\mathrm{p}}{R^2},
\end{equation}
where $G$ is the gravitational constant.  The maximally allowed proto-planetesimal radius before the gravitational force on the dust particles becomes comparable to the drag force is then
\begin{equation}
\label{eq:radgrav}
R_\mathrm{grav} \approx 250\, \mathrm{m}.
\end{equation}

Note also that the gravity of a proto-planetesimal with $R \lesssim R_\mathrm{grav}$ does not affect the disk gas because the escape velocity from its surface is much smaller than the local thermal speed of the gas molecules.

In assumption (a), we found that the minimally required radius of a proto-planetesimal which induces a hydrodynamical flow grows with heliocentric distance. There must consequently be a distance when the radius exceeds the above specified $R_\mathrm{grav}$. Setting $R_{\min} = R_\mathrm{grav}$ in Eqs.~(\ref{eq:rminid}) \& (\ref{eq:rminim}), this occurs at
\begin{align}
\label{eq:rgravd}
r_{\mathrm{grav},\desch} &= 14\, \mathrm{AU}; \\
\label{eq:rgravm}
r_{\mathrm{grav},\mmsn} &= 8\, \mathrm{AU},
\end{align}
for the Desch and MMSN models, respectively. Beyond these heliocentric distances, a proto-planetesimal cannot fulfill both prerequisites of being large enough to induce a hydrodynamical flow and small enough for gravity to be negligible, and our model is no longer valid.
\section{Hydrodynamical effects on collisions between proto-planetesimals and dust grains}
\label{effects}
In order to investigate the change of the effective sweep-up cross section $\sigma_{\mathrm{eff}}$ and the impact velocity $v_\mathrm{imp}$ due to the drag in a hydrodynamical flow pattern, we calculate the trajectories of dust particles from the equation of motion,
\begin{equation}
\dot{\vec{v}} = -\frac{(\vec{v} - \vec{v}_\mathrm{g})}{\tstop},
\label{eq:eom}
\end{equation}
where the dot denotes a time derivative, $\vec{v}$ is the dust particle's velocity, $\vec{v}_\mathrm{g}$ is the velocity of the gas flow (Eq.~\ref{eq:hydro}), $t_\mathrm{s} = |\vec{p}| / F_\mathrm{D}$ is the stopping time of the dust particle, and $|\vec{p}| = m|\vec{v}|$ is the dust particle momentum. For example, in the Epstein regime, when the particle radius $a < 9\lambda/4$, the stopping time is given by $t_\mathrm{s,Ep} = \rho_\mathrm{p} a / \rho_\mathrm{g} \bar{u}_\mathrm{th}$. This quantity can be regarded as a particle property parameterizing the effects which contribute to the drag of this particle in a specified medium. Hence, the following work can be applied to real protoplanetary disks if stopping times of dust particles are known. Theoretical stopping times for particles in different drag regimes are summarized in \citet{WeidAero}.
\subsection{Qualitative aspects of the flow pattern and sweep-up cross section}
\label{qualitative}
\begin{figure}[ht]
  \begin{center}
  \resizebox{1.0\hsize}{!}{\includegraphics{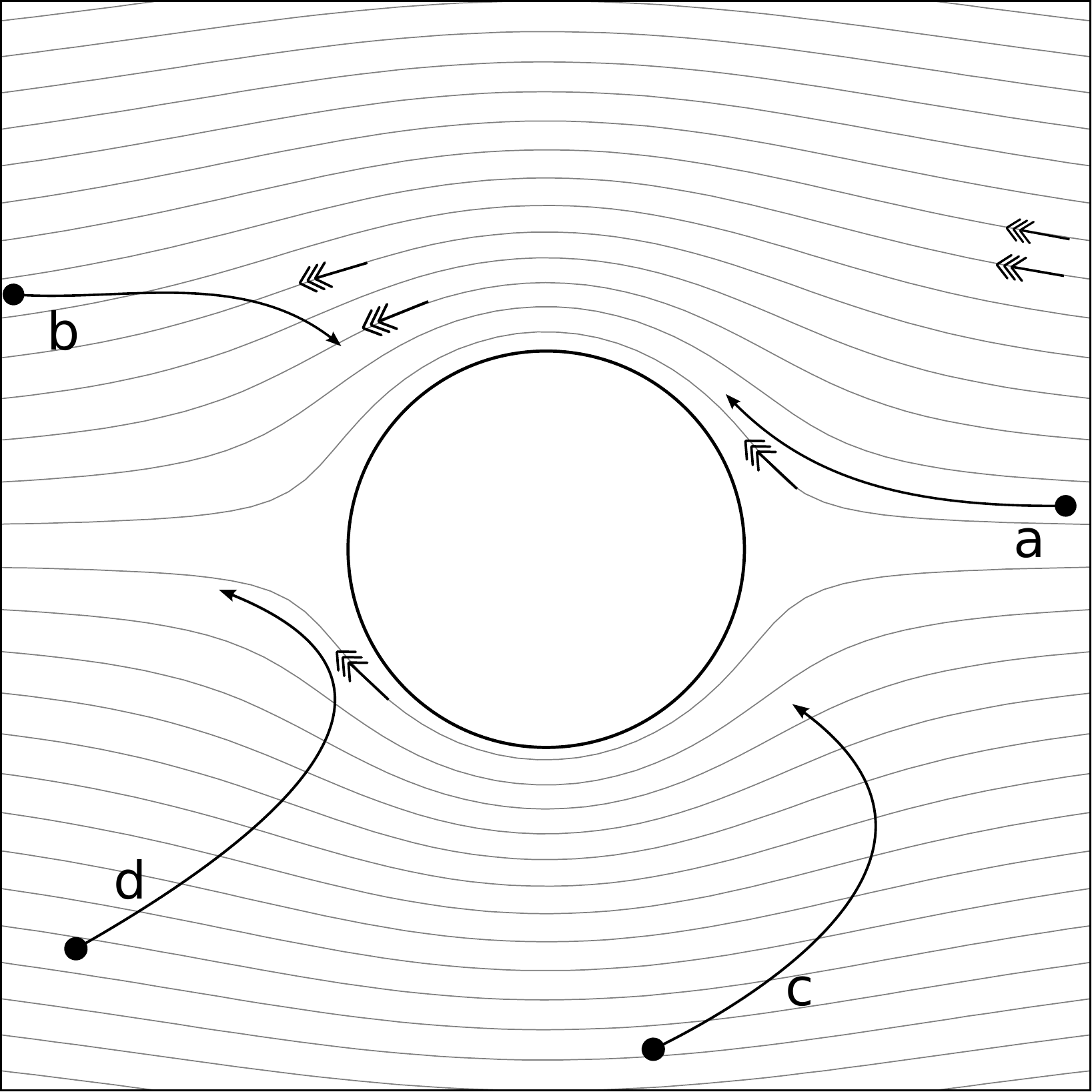}}
  \caption{Trajectories of dust particles resulting from their interaction with the gas flow are indicated by the solid arrows. The gray lines represents stream lines of the gas flow around the proto-planetesimal, as calculated from Eq.~(\ref{eq:hydro}). The feathered arrows indicate the direction of gas flow.  See the text for a discussion of scenarios (a) -- (d).}
  \label{fig:Flowpattern}
  \end{center}
\end{figure}

Far upstream and downstream of a spherical proto-planetesimal, the streamlines of the chosen flow pattern Eq.~(\ref{eq:hydro}) are parallel. Meanwhile, near the proto-planetesimal, the stream lines diverge upstream and converge again downstream, creating a flow pattern that is rotationally symmetric about the flow axis. If dust particles approach the proto-planetesimal in this flow pattern from all directions -- which could be possible if they are stirred up by nearby turbulent eddies and acquire randomly orientated velocities, or if the headwind velocity is modified by a large-scale turbulent eddy -- then they should experience the following scenarios, all of which are depicted in Fig.~\ref{fig:Flowpattern}:
\begin{enumerate}[{arrow} a)]
\item Dust particles that approach the proto-planetesimal from upstream and parallel to the flow axis will be deflected away from the proto-planetesimal due to the diverging flow pattern. In particular, particles with a large impact parameter that would collide with the proto-planetesimal in the absence of the flow pattern can be deflected such that they pass over the rim of the proto-planetesimal and miss it. The collisional cross section of the proto-planetesimal is then reduced relative to its geometrical cross section.  Furthermore, the magnitude of the deflection will depend on the dust particle stopping time: The larger the stopping time, the smaller the deflection.  Lastly, due to the rotational symmetry of the flow pattern, the collisional cross section will be circular. 

\item Particles that approach the proto-planetesimal from downstream and parallel to the flow axis do so under a converging gas flow. Even particles that lie outside the geometrical cross section of the proto-planetesimal can then be pushed onto a collision course by the gas flow. A collision, however, only occurs if the particles have sufficient inertia such that the gas flow cannot reverse their direction of motion (\ie\ a large stopping time). For particles that approach from downstream and collide with the proto-planetesimal, the collisional cross section must then be larger than the geometrical cross section, but will remain circular due to rotational symmetry.

\item Particles that approach the proto-planetesimal at an angle $\theta$ to the flow axis experience the flow pattern as a cross wind. Particles not initially on a collision course can be deflected such that they now collide with the proto-planetesimal.  In this scenario, the cross wind introduces an additional dependence on $\lvert\sin{\theta}\rvert$ to the collisional cross section.  For a spherical proto-planetesimal, this causes a small stretching of the collisional cross section in the upstream direction, resulting in a top-down symmetric oval. The slight breaking of left-right symmetry occurs because the flow pattern also depends on $\theta$ (Eq.~\ref{eq:hydro}).

\item In contrast, particles that approach the proto-planetesimal under an angle to the flow axis, but that were originally on a collision course, can be deflected such that they now miss the proto-planetesimal.
\end{enumerate}
To test for the above qualitative features of the dust trajectories, we ran a set of example simulations (see Sect.~\ref{setup}), and found that all cases occur as described. The reduction of the proto-planetesimal collisional cross section for particles that approach from upstream (arrow (a) in Fig.~\ref{fig:Flowpattern}) was found to be the most significant, while the enhancement (or distortion) of the cross section for particles that approach from downstream (or under an angle) is effectively negligible.

We expect particles that approach from upstream to be the most frequently occurring situation in a sweep-up scenario, and thus in the following we focus only on the case of upstream particles travelling parallel to the flow axis. 
\subsection{Description of simulations and initial conditions}
\label{setup}
\begin{table*}[htb]
\caption{Simulation parameter space. The range of values for each parameter is determined with regards to Sect.~\ref{assumptions}.}
\centering
\begin{tabular}{c c c c}
\hline\hline
Parameter  & Definition                & Range                         & As restricted by  \\
\hline
$R$        & proto-planetesimal radius & $1 - 150~ \mathrm{m}$         & gas mean free path \& neglect of gravity \\
$v_\infty$ & headwind velocity         & $15 - 40~ \mathrm{m\,s^{-1}}$ & protoplanetary disk model\\
$t_\mathrm{s}$ & dust particle stopping time & $10^{-1} - 10^4~ \mathrm{s}$ & size of perturbed gas region \\
\hline
\end{tabular}
\label{table:1}
\end{table*}
In total, we performed 16000 simulations that numerically integrate Eq.~(\ref{eq:eom}) for the trajectories of dust particles in the flow pattern around a proto-planetesimal using a 4\textsuperscript{th}-order Runge-Kutta based, backwards Euler method with adaptive stepping. Each simulation uses a different combination of proto-planetesimal radii, headwind velocities, and particle stopping times, constrained in Sect.~\ref{assumptions} and summarized in Table~\ref{table:1}.

The calculation of the headwind velocity $v_\infty$ in Sect.~\ref{assumptions} only accounts for the sub-Keplerian gas velocity as the origin of a relative motion between disk gas and proto-planetesimals. The total headwind velocity should also include, for example, disk turbulence, and for the sake of generality we vary the headwind velocities between $v_\infty \in [15, 60]\, \mathrm{m\, s^{-1}}$ (\cf\ the headwind velocities calculated in \citealt{WeidAero}).

In our simulations of particle trajectories, we assume the particles are initially at rest with respect to the gas, and thus we set the initial velocity of the dust particles equal to the headwind velocity.  Although dust particles that react significantly to the disturbed gas flow around the proto-planetesimal consequently must have a relatively small stopping time, and are thus essentially comoving with the gas, only dust particles with $\tstop = 0$ will always remain perfectly at rest with respect to the gas.

The dust particles start at a distance $H\cdot R$ upstream from the proto-planetesimal, where the parameter $1 < H \leq 150$ is a multiple of the proto-planetesimal radius. The larger the value of $H$, the closer the gas is to a uniform flow pattern, and the longer the particles will remain at rest with respect to the gas.

\begin{figure}[ht]
  \begin{center}
  \resizebox{1.0\hsize}{!}{\includegraphics[clip=true,trim=25 20 0 0]{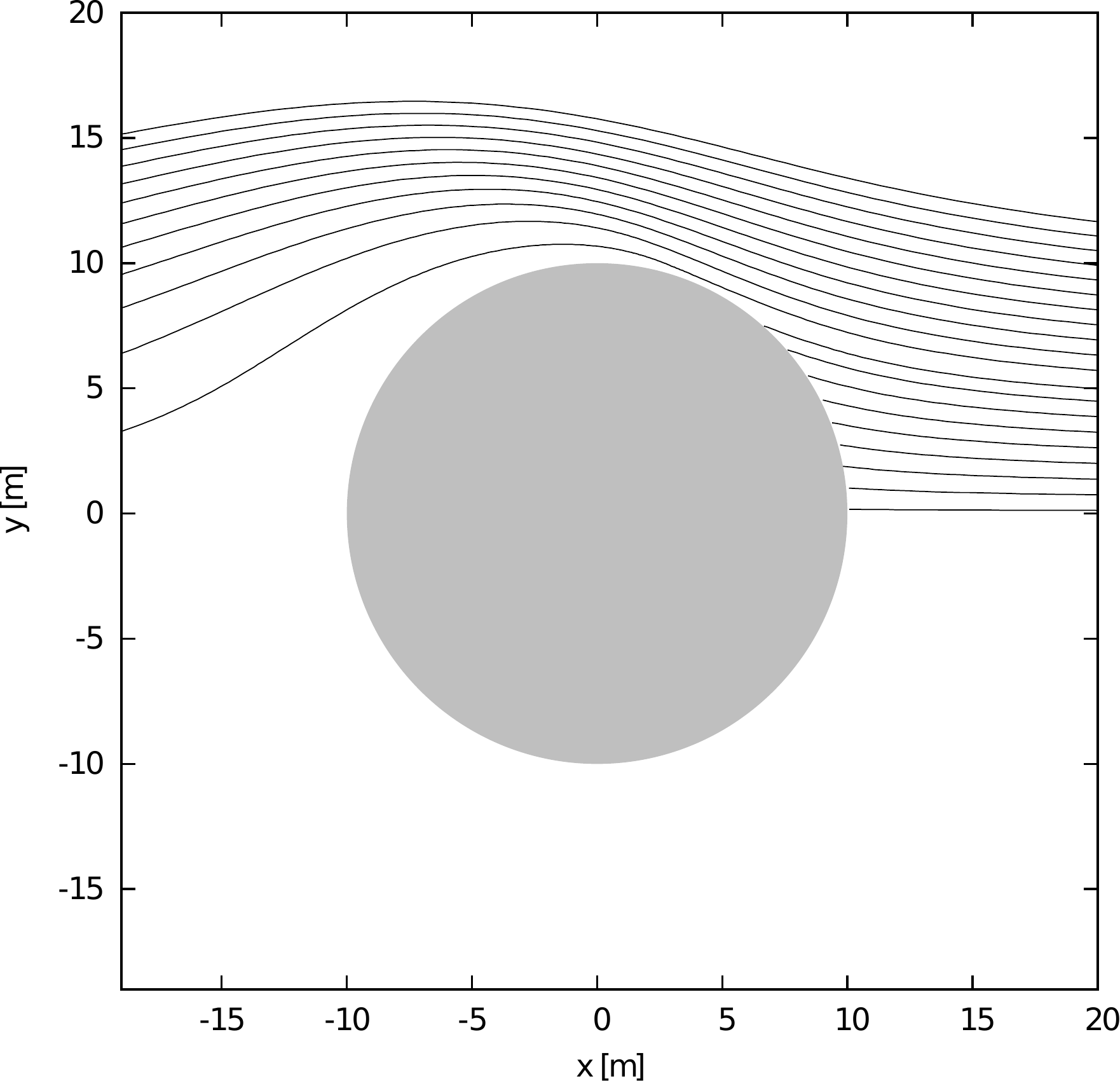}}
  \caption{The numerically calculated trajectories of dust particles with different impact parameters in the viscous laminar flow around a spherical proto-planetesimal are indicated by solid lines. In this representative example, the proto-planetesimal radius is $10 ~\mathrm{m}$, the headwind velocity of the gas is $20~\mathrm{m\, s^{-1}}$, and the particle stopping time is $1~\mathrm{s}$. The dust particles enter the plot from the right, and the axes have units of meters.}
  \label{fig:Mercedes}
  \end{center}
\end{figure}

To investigate which impact parameters lead to collisions, we vary the initial impact parameter $p$ in steps of $10^{-4}R$ between zero and $0.01 R$, and then by steps of $0.01 R$ up to the planetesimal radius $R$. Fig.~\ref{fig:Mercedes} shows a representative example of trajectories for dust particles with different impact parameters.

While particles with small impact parameters collide with the proto-planetesimal, beyond a certain value of $p = p_\mathrm{max}$ ($< R$) the particle will be deflected around the proto-planetesimal, and impacts cease to occur. The maximum impact parameter, $p_\mathrm{max}$, that leads to a collision is related to the effective sweep-up cross section $\sigma_{\mathrm{eff}}$ by:
\begin{equation}
\label{eq:sigmaeff}
\sigma_{\mathrm{eff}} = \pi p_{\mathrm{max}}^2.
\end{equation}
\subsection{The influence of the flow pattern on the sweep-up cross section}
\label{sweepupcross}
For the entire set of 16000 simulations, we find the outcome depends only on the dimensionless parameter:
\begin{equation}
\label{eq:parameterx}
x:=\frac{R}{v_\infty \tstop}.
\end{equation}
This parameter $x$ is the ratio of the hydrodynamical time scale for flow past the proto-planetesimal, $t_\mathrm{p} = R/v_\infty$, and the particle stopping time, $\tstop$.  Note that  $x$ is equal to the reciprocal of the dimensionless stopping time (\cf\ \citetalias{Sek}). We prefer to use $x$ here because it characterises the strength of the deflection by the flow for a particle with a given $\tstop$. As $x$ increases, so does the particle deflection, eventually reaching a point where the particle no longer collides with the proto-planetesimal and the effective sweep-up cross section goes to zero. Conversely, as $x \rightarrow 0$, the influence of the gas flow on the particle becomes insignificant, and the resulting effective cross section reduces to the geometrical cross section.

When $x>0.8$, we observe that $\sigma_\mathrm{eff}/\sigma_\mathrm{geom} < 10^{-4}$, and we therefore consider it to be zero.  As such, for $x>0.8$, the deflection of dust particles is substantial enough to entirely prevent impacts onto the proto-planetesimal, and the particles are instead advected around it.  That collisions cease for $x < 1$ means slightly less than one stopping time is required for the flow to deflect a particle around the proto-planetesimal.

\begin{figure}[ht]
  \begin{center}
  \resizebox{1.0\hsize}{!}{\includegraphics[angle=-90]{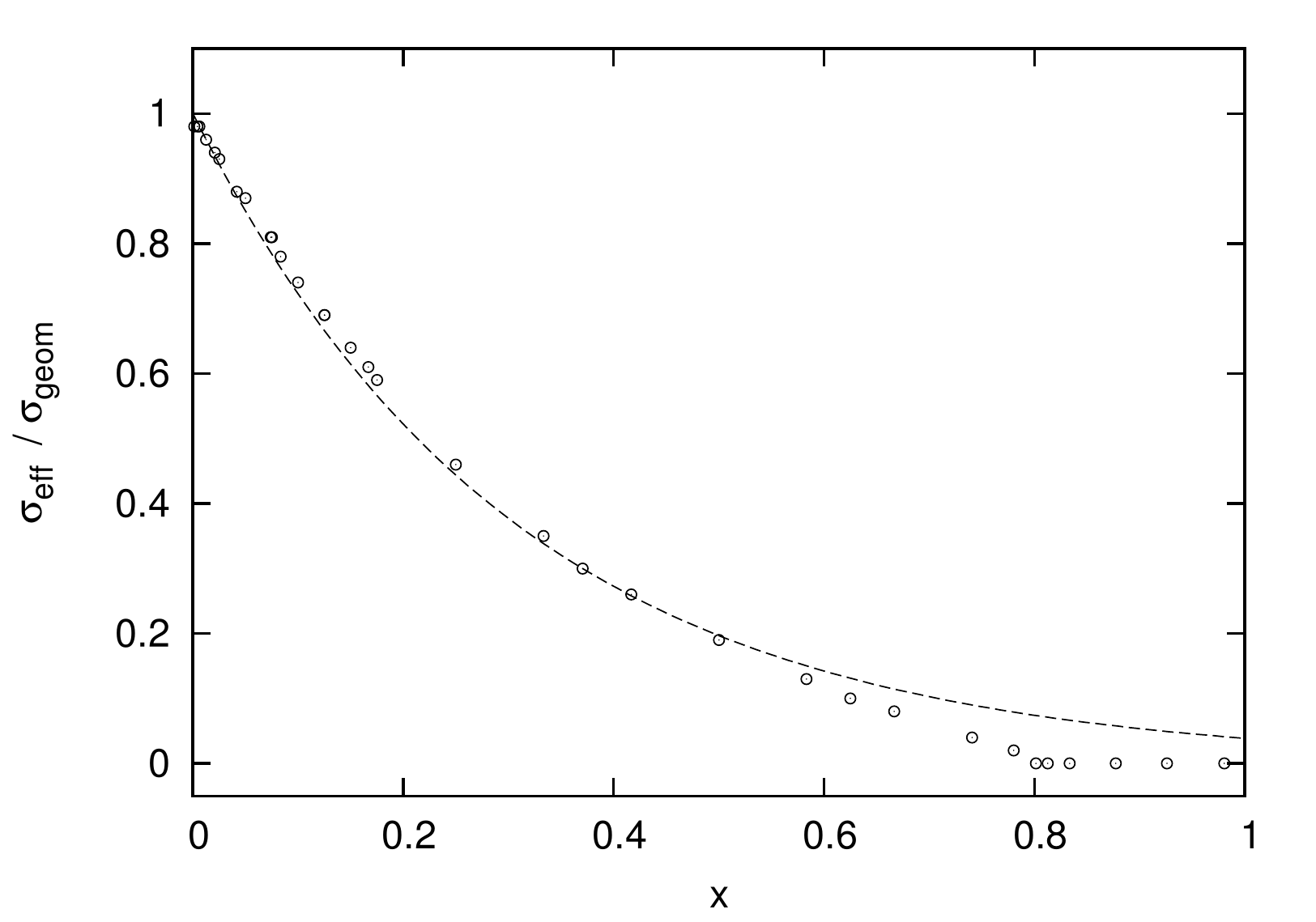}}
  \caption{Decreasing effective sweep-up cross section as a function of the parameter $x$ for $H = 10$. The open circles represent simulation data, and the dashed line corresponds to the best-fitting function Eq.~(\ref{eq:sigma}). Above $x > x_\mathrm{cutoff}$, the simulations show that particles no longer impact the proto-planetesimal. In this regime, the best-fitting function overestimates the sweep-up cross section.}
  \label{fig:worstSigma}
  \end{center}
\end{figure}

Fig.~\ref{fig:worstSigma} demonstrates how the effective sweep-up cross section varies with the parameter $x$. The open points are taken from our simulations, while the dashed line plots the function which best-fits the data points. This function is given by
\begin{equation}
\frac{\sigma_\mathrm{eff}}{\sigma_\mathrm{geom}} = \exp\left( -D_\sigma(H)\, x \right),
\label{eq:sigma}
\end{equation}
where $\sigma_\mathrm{geom} = \pi R^2$ is the geometrical cross section. The function $D_\sigma(H)$ accounts for the dependence of the effective cross section (\ie\ the amount of deflection) on the starting distance $H$ from the surface of the proto-planetesimal.

Although Eq.~(\ref{eq:sigma}) typically fits the simulation results to within $10\%$ difference, it does not reproduce the effect that particles no longer impact the proto-planetesimal for $x > 0.8 := x_\mathrm{cutoff}$. We therefore suggest the following form for the best-fit function:
\begin{equation}
  \label{eq:stepfunc}
  \frac{\sigma_\mathrm{eff}}{\sigma_\mathrm{geom}} = \begin{cases} ~ \exp\left( -D_\sigma(H)\, x \right) & x < x_\mathrm{cutoff}; \\
  ~0 & x \geq x_\mathrm{cutoff}. \end{cases}
\end{equation}

If a dust particle starts very close to the proto-planetesimal surface, \ie\ $H \gtrsim\, 1$, the flow does not have a significant opportunity to deflect the particle and the sweep-up cross section is then very close to the geometrical cross section. Comparing simulations with different values of $H$, we find:
\begin{equation}
D_\sigma(H) = 4.00 - 4.75\cdot \exp\left(-\frac{H}{5.38} \right).
\label{eq:DvH}
\end{equation}
The maximum deviation of Eq.~(\ref{eq:DvH}) from the simulation results is $10\%$, although the typical deviation is only $\sim 2\%$.

For $H > 100$, the function $D_\sigma(H) \simeq 4$. This behaviour is physical: In Sect.~\ref{assumptions} we calculated that the viscous, laminar flow pattern reaches a gas velocity of $98\%$ the upstream velocity $v_\infty$ at a distance of $100\, R$. That the function $D_\sigma(H)$ is constant beyond $H=100$ therefore originates from the gas velocity being effectively constant in this regime.

In a protoplanetary disk, it is reasonable to assume that the collision timescale for small dust is long enough that the particles have spent several stopping times in the gas flow, and they are at rest with respect to the gas as they approach the proto-planetesimal. Thus, one can safely choose the asymptotic value $D_\sigma(H) = 4$ when applying Eq.~(\ref{eq:sigma}) to a sweep-up growth scenario.

If one wants to consider the influence of turbulence on particle velocities, then one might need to consider small values of of $H$ and $D_\sigma(H) < 4$.  However, this means the particle is not initially at rest with respect to the gas flow, nor travelling parallel to the flow axis, and this is beyond the scope of the current study.
\subsection{The influence of the flow pattern on the impact velocity}
\label{impvel}
Here, we summarize the effects that we observe for the flow pattern on the measured impact velocity $v_\mathrm{imp} = |\vec{v}_\mathrm{imp}|$ of dust particles onto the proto-planetesimal. From Eq.~(\ref{eq:hydro}), the gas velocity in the perturbed region steadily decreases as a particle approaches the proto-planetesimal surface. Directly at the surface, the gas velocity is zero. The reduced gas velocity exerts a drag force on approaching dust particles, thereby reducing their impact velocity.

In our simulations, we find the impact velocity to be virtually independent of the impact parameter -- under the condition that the chosen impact parameter leads to a collision at all. In the following, we therefore take the average of the impact velocities, weighted by the impact parameter $p$, over all impact parameters that lead to collisions and refer to it as \emph{the} impact velocity $v_{\mathrm{imp}}$.

\begin{figure}[ht]
  \begin{center}
  \resizebox{1.0\hsize}{!}{\includegraphics[angle=-90]{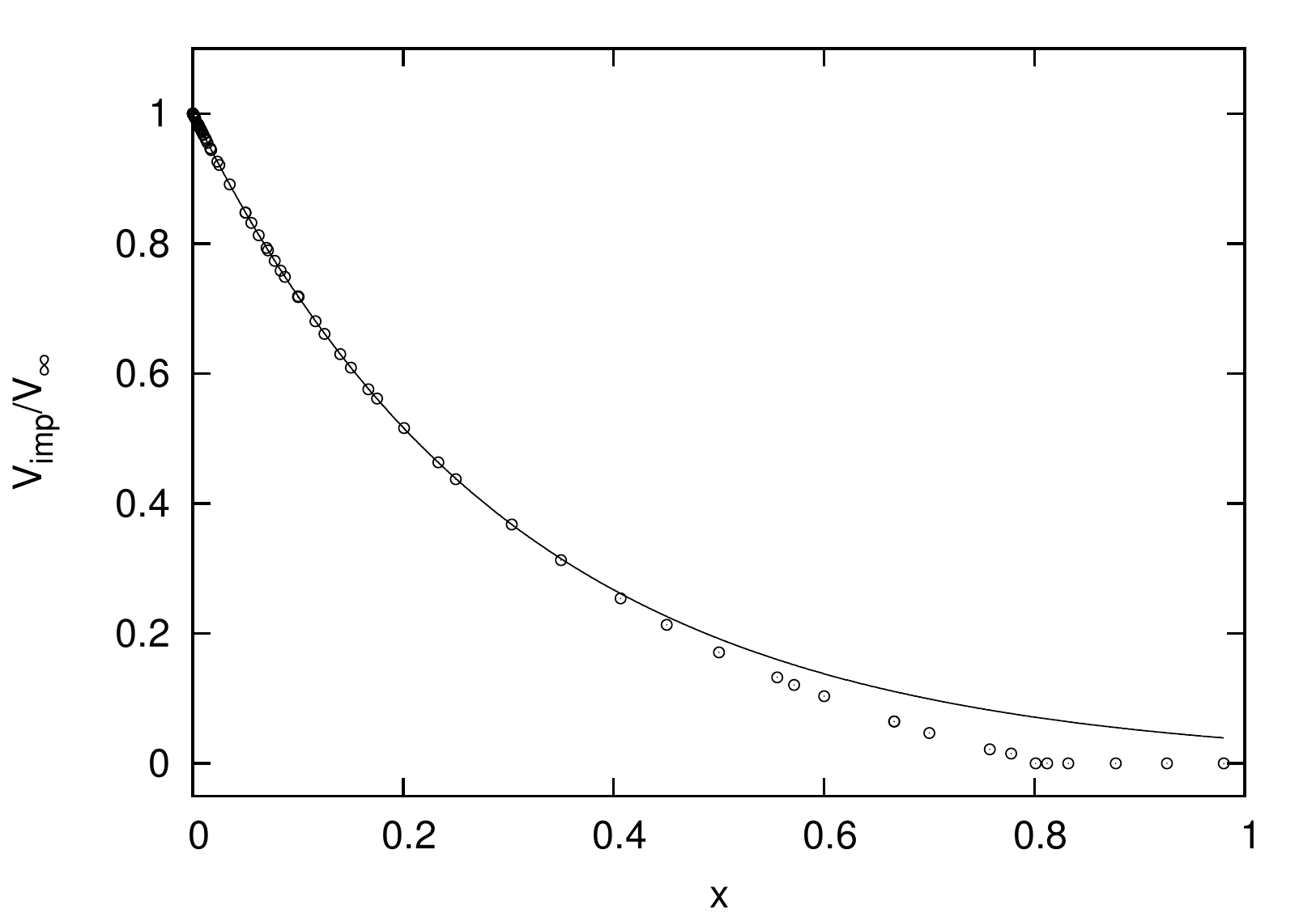}}
  \caption{Decreasing impact velocity of dust particles onto a proto-planetesimal as a function of the parameter $x$. The open circles represent simulation data, while the solid line corresponds to the best-fitting function Eq.~(\ref{eq:vimp}).}
  \label{fig:Vimpaver}
  \end{center}
\end{figure}

Fig.\ (\ref{fig:Vimpaver}) demonstrates how the impact velocity of dust particles varies with $x$. The open circles are taken from our simulations and the solid line depicts the best fitting function:
\begin{equation}
  \frac {v_\mathrm{imp}}{v_\infty} = \exp\left(-D_v x \right);~ D_v = 3.3.
\label{eq:vimp}
\end{equation}
Eq.~(\ref{eq:vimp}) describes that particles with small $x$ will impact the proto-planetesimal with velocity equal to $v_\infty$. The impact velocity decreases with increasing $x$ because the drag that brakes the particles also increases with $x$.

Eq.~(\ref{eq:vimp}) fits the simulation results to better than $10\%$ and, as with Eq.~(\ref{eq:sigma}), Eq.~(\ref{eq:vimp}) is only applicable for $x < x_\mathrm{cutoff}$ because an impact velocity is ill-defined for particles that do not impact the proto-planetesimal.

\citet{Schraep} find that large dust particles are prone to erosion by the impact of smaller particles in a process that has become known as monomeric erosion. More specifically, they find the mass loss of the target scales linearly with the impact velocity of the monomers, and that the process can result in losses of up to $\sim 10\, \times$ the mass of the impactor.

We find that a flow around proto-planetesimals reduces the impact velocity of small dust particles (\ie\ large $x$, all else being equal), and thus will have the consequence of reducing the efficiency of monomeric erosion. In the extreme limit of $x > x_\mathrm{cutoff}$, the flow entirely prevents impacts and thereby also monomeric erosion. Thus, if a proto-planetesimal grows large enough that a hydrodynamical gas flow develops around it, it can become partially shielded against erosive high velocity impacts.
\subsection{Dependence of the results on the Reynolds number}
\label{sub:realitycheck}
\begin{figure*}[ht]
  \begin{center}
  \resizebox{1.0\hsize}{!}{\includegraphics{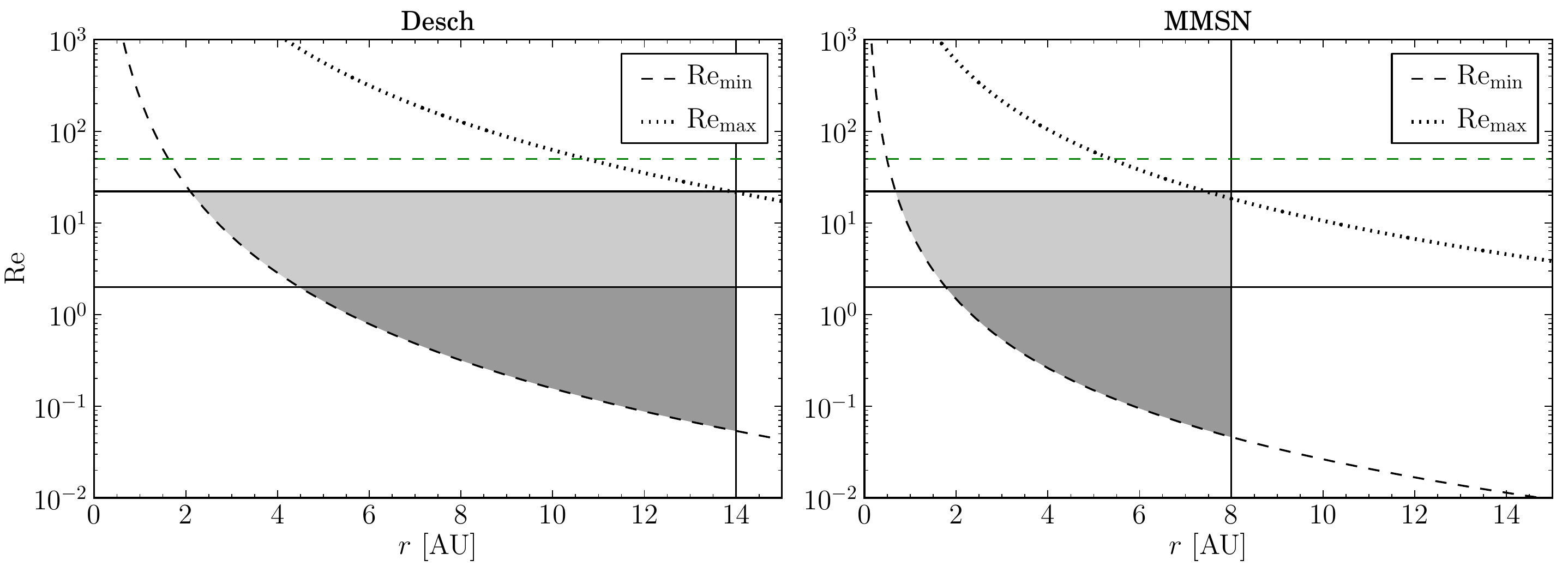}}
  \caption{Reynolds number as a function of heliocentric distance $r$ for the Desch and MMSN disk models.  The $\ReN_\mathrm{min}$ (dashed) line corresponds to a proto-planetesimal radius and headwind velocity of $(R, v_\infty)  = (1\, \mathrm{m}, 15\, \mathrm{m\, s^{-1}})$, while the $\ReN_\mathrm{max}$ (dotted) line corresponds to $(150\, \mathrm{m}, 40\, \mathrm{m\, s^{-1}})$.  The solid horizontal lines denote Reynolds numbers of 2 and 22, while the solid vertical line indicates the critical gravitational heliocentric distance for each disk model (Eqs.\ \ref{eq:rgravd} \& \ref{eq:rgravm}).  The successively darker shading distinguishes the regions of our parameter space that fall below Reynolds numbers of 22 and 2, respectively.}
  \label{fig:reynoldsnumber}
  \end{center}
\end{figure*}

From Eqs.\ (\ref{eq:viscid}), (\ref{eq:viscim}), and the range of values for $R$ and $v_\infty$ in Table \ref{table:1}, one can calculate the minimal and maximal values of the Reynolds numbers for our parameter space:
\begin{align}
\label{eq:minmaxRe}
\ReN_\mathrm{max} &= 2 R_\mathrm{max} v_{\infty, \mathrm{max}} / \nu (r);\nonumber \\
\ReN_\mathrm{min} &= 2 R_\mathrm{min} v_{\infty, \mathrm{min}} / \nu (r).\nonumber
\end{align}
The resulting range of values, constrained by the assumptions of Sect.\ \ref{assumptions}, are shown in Fig.\ \ref{fig:reynoldsnumber}.  For example, at $r = 5$ AU, in the Desch (MMSN) model with a headwind velocity of $24\, \mathrm{m\, s^{-1}}$ ($20\, \mathrm{m\, s^{-1}}$), requiring $\ReN \leq 22$ restricts our results to proto-planetesimal radii $R \leq\, 9.8\, \mathrm{m}$ ($\leq\, 110.7\, \mathrm{m}$).

Eq.\ \ref{eq:hydro} assumes $\ReN \ll 1$, and is therefore strictly valid only in this regime.  Meanwhile, experimental results affirm that the flow streamlines upwind of a sphere remain remarkably similar to Eq.\ (\ref{eq:hydro}) for Reynolds numbers $\ReN \lesssim 22$ (\eg\ \citealt{Taneda56, vandyke1982} and references therein).  At larger Reynolds numbers however, the flow pattern downstream of the sphere qualitatively changes as the flow separates and vortices begin to form.

As we are only concerned with the sweep-up of dust particles on the upstream side of the proto-planetesimal, the application of Eq.\ (\ref{eq:hydro}) is thus expected to accurately describe the deflection of particles even when Re $\centernot\ll 1$.

In an attempt to quantify this expectation, we examined two additional analytical approximations for viscous flow past a sphere, described in detail in \citet{vandyke1964}, and which explicitly depend on the Reynolds number.  The first approximation, due to Oseen, accounts for the convective terms of the full Navier-Stokes equations (which Eq.\ \ref{eq:hydro} ignores) via linearisation.  However, this approximation suffers from the issue that the velocity at the surface of the sphere depends on a term $\mathcal{O}(\ReN)$.  In contrast, Eq.\ (\ref{eq:hydro}) correctly predicts zero velocity at the surface of the sphere.

The second, which we refer to here as ``Proudman and Pearson's two-term approximation'' (\citealt{pp57}; ``PP's approximation'' for short), is derived with the region near the sphere in mind, and preserves a zero velocity at the surface.  Indeed, in the immediate vicinity of the sphere, this approximation does a remarkable job of matching experiment, even for $\ReN > 22$ \citep[p.\ 160]{vandyke1964}.  The trade-off, however,  is that the velocity far from the sphere can significantly exceed $v_\infty$ when $\ReN > 1$ and, in general, does not return to a uniform flow (as do Eq.\ \ref{eq:hydro} and Oseen's approximation).  Consequently, particles can be accelerated beyond $v_\infty$, impacting the proto-planetesimal with $v_\mathrm{imp}/v_\infty > 1$.

Although both approximations are strictly valid only for small Reynolds numbers, because of their explicit dependence on $\ReN$, we apply them here to provide some intuition on the behaviour of particle accretion when $\ReN > 1$.  Furthermore, as our results strongly depend on the conditions in the immediate vicinity of the proto-planetesimal, and Oseen's approximation does not recover a zero velocity at the surface of the proto-planetesimal, we choose to henceforth discuss only PP's approximation.

Relative to PP's approximation, we expect simulations applying Eq.\ (\ref{eq:hydro}) to progressively overestimate the deflection of dust particles as $\ReN$ increases.  Indeed, additional simulations run using PP's approximation demonstrate that $x_\mathrm{cutoff}$ increases linearly with $\ReN$.  In other words, as the relative importance of inertial forces increase, all else being equal, a successively smaller stopping time is required for a particle to be deflected around the proto-planetesimal.

Similarly, the additional simulations show that $D_\sigma(H)$ and $D_v$ decrease with increasing $\ReN$ before asymptoting to small but non-zero [$\mathcal{O}(10^{-1})$] values for $\ReN \gtrsim 15$.  This manifests as an upward shift of the curves in Figs.\ \ref{fig:worstSigma} \& \ref{fig:Vimpaver} and, as with $x_\mathrm{cutoff}$, this illustrates that the amount of particle deflection decreases with increasing $\ReN$.

In regards to the accuracy of using Eq.\ (\ref{eq:hydro}) when $\ReN \centernot\ll 1$, we observe that $x_\mathrm{cutoff}$, $D_\sigma(H)$, and $D_v$ match the results from PP's approximation for $\ReN \lesssim 0.1$.  Moreover, these variables agree with PP's approximation to within 50\% for $\ReN \leq 1$, and within a factor of two for $\ReN \leq 2$.  Beyond $\ReN = 2$, notwithstanding the variation of $x_\mathrm{cutoff}$, $D_\sigma(H)$, and $D_v$, the results applying Eq.\ (\ref{eq:hydro}) remain qualitatively correct with respect to PP's approximation.  Thus, Fig.\ 5 depicts where in our parameter space the results of Stokes' and PP's approximations agree to within a factor of 2 (dark grey), and where these approximations still qualitatively agree (light gray).

That said, keep in mind that the additional approximations considered here are derived under the assumption of small $\ReN$, and thus the above trends are at best estimates.  It should also be noted that the effects of $\ReN$ on $x_\mathrm{cutoff}$, $D_\sigma(H)$, and $D_v$ can at least be partially explained by the locally enhanced particle velocities (relative to $v_\infty$) observed when using PP's approximation.  Direct numerical simulations of subsonic, laminar flow past a proto-planetesimal are needed to verify these results, and this is left to future work.\footnote{For direct numerical simulations of particle accretion in fully turbulent flow see, for \eg, \citealt{mitraetal2013}.}.

\section{Hydrodynamical effects on the reaccretion of collisional debris}
\label{reaccretion}
If a dust aggregate collides with a proto-planetesimal and the collision velocity is large enough, the aggregate will be disrupted into fragments. Debris of the particle can then re-enter the gas flow around the proto-planetesimal. In the literature, the question has been raised whether the gas flow can return such fragments to the surface of the proto-planetesimal, and thereby increase the efficiency of sweep-up growth \citep{WurmBlumColwell2001, WurmKraus, Sek, SekiyaTakeda2005}.

In this section, we attempt to address the escape of collisional debris from an impact site by launching dust particles directly from the surface of the proto-planetesimal into the flow pattern with random directions.
\subsection{On the free molecular flow and straight trajectories}
\label{reaccretefree}
\citet{Sek} find that the gas can return collisional debris to the proto-planetesimal surface in the free molecular flow. The term free molecular flow describes the situation where the proto-planetesimal radius is comparable to or smaller than the mean free path of the gas. The possibility of averaging over the random thermal motion of the gas particles is then no longer available, and the fluid has instead to be described by the motion of individual particles.

Even in free molecular flow, the transformation into the rest frame of the proto-planetesimal induces an apparent macroscopic motion of the gas, namely the headwind. \citetalias{Sek} assume the gas motion can then be described by straight, parallel streamlines that intersect with the proto-planetesimal.

However, the thermal motion of the gas will be superimposed on the ordered headwind. From the temperature profile Eq.~(\ref{eq:temp}), the mean thermal speed of gas molecules in the disk is then:
\begin{equation}
\label{eq:thermalspeed}
\bar{u}_{\mathrm{th}}(r) \simeq 960\, \rau^{-1/4}\, \mathrm{m\, s^{-1}}.
\end{equation}
At 1 AU, the random thermal motion outweighs the headwind velocity of $v_\infty \approx 24\,\mathrm{m\, s^{-1}}$ ($20\,\mathrm{m\, s^{-1}}$) by a factor of 40 (48), and by a factor of $\gtrsim 20$ (28) within the inner 14 AU (8 AU) of the disk for the \citeauthor{Desch} (MMSN) model. Clearly, the ordered macroscopic motion of the headwind will disappear under the random thermal motion of the free molecular flow, and thus parallel stream lines are not physically applicable in this context.

However, if we launch dust particles from the surface of the proto-planetesimal into a flow pattern of parallel stream lines, then we do reproduce the result of \citetalias{Sek}, namely that parallel streamlines lead to reaccretion of collisional debris on the upstream side of the proto-planetesimal.
\subsection{Reaccretion in a viscous laminar flow}
\label{reaccretelaminar}
We also launched collisional debris from the surface of the proto-planetesimal into the laminar flow pattern of Eq.~(\ref{eq:hydro}) with multiple ejection velocities and angles in the range $0 \leq \alpha \leq \pi$.

Independent of the ratio between headwind velocity and the ejection velocity of the collisional debris, we find virtually no reaccretion, in agreement with the findings of \citetalias{Sek}. In our simulations however, we observe a maximal fraction of $\sim\! 10^{-3}$ of particles return to the proto-planetesimal surface.  These re-impacts only occur for angles nearly tangential to the proto-planetesimal surface, and it is our opinion that they are the result of limited numerical precision and thus, physically, no reaccretion occurs.

\begin{figure}[ht]
  \begin{center}
  \resizebox{1.0\hsize}{!}{\includegraphics{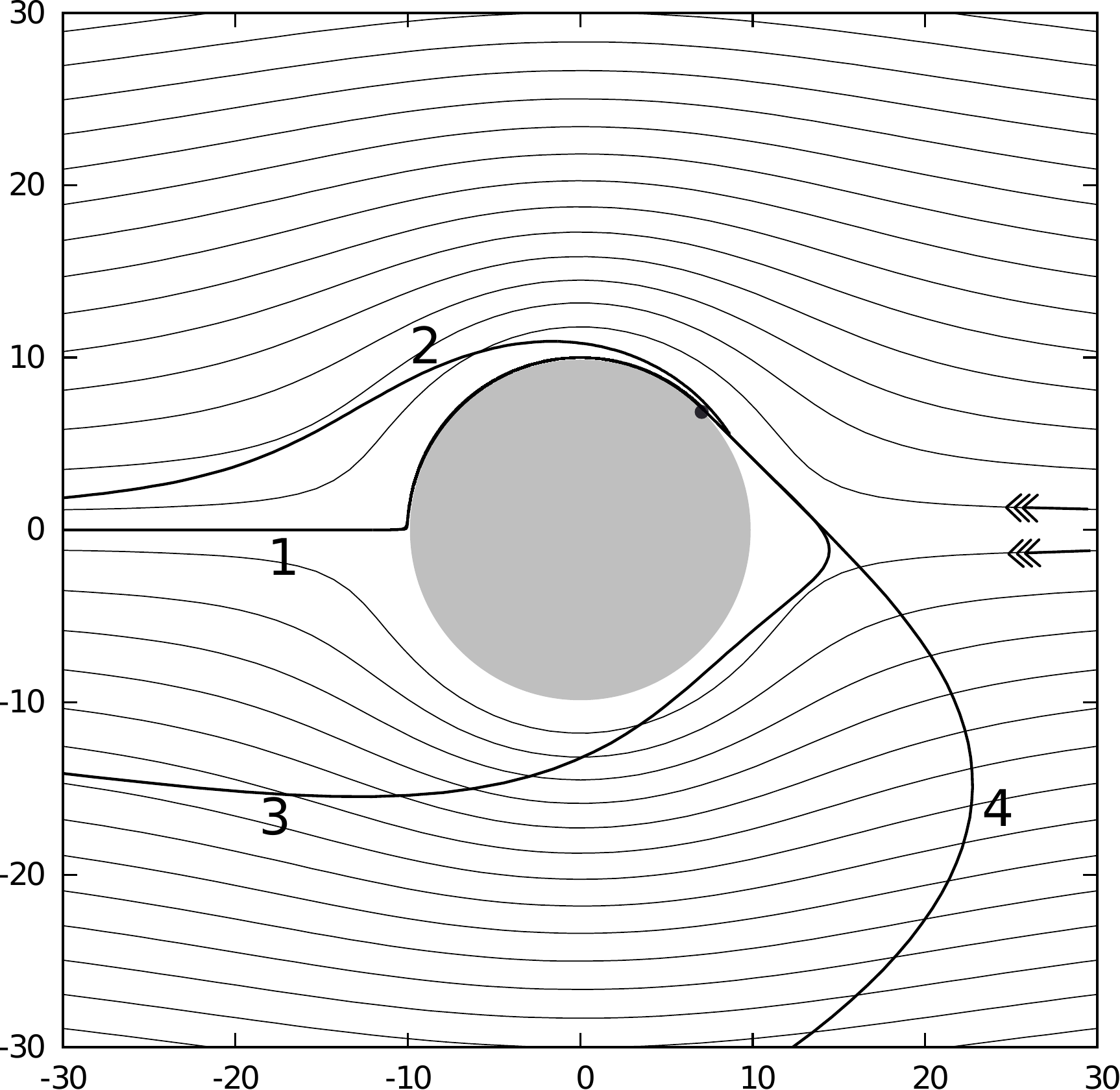}}
  \caption{Trajectories of collisional debris in a laminar flow. In this representative example, the headwind velocity is $20\ \mathrm{m\, s^{-1}}$, and the axes have units of meters. All dust particles begin with the same location, direction $\alpha$, and velocity ($20\, \mathrm{m\, s^{-1}}$), indicated by a dot on the proto-planetesimal surface. The stopping times of the dust particles are, for trajectories 1 -- 4, $\tstop = 0.1$, $1$, $8$, and $40\, \mathrm{s}$, respectively. See the text for a discussion of the different outcomes.}
  \label{fig:LamMercedes}
  \end{center}
\end{figure}

Fig.~\ref{fig:LamMercedes} shows a representative example of the possible trajectories for collisional debris. Plotted are the trajectories of four dust particles that leave the proto-planetesimal surface with the same initial location, direction $\alpha$, and ejection velocity ($2\, \mathrm{m\, s^{-1}}$), but with different stopping times.

Particles with small stopping times only escape into the fluid layers just above the proto-planetesimal surface before being deflected (trajectories 1 and 2). Since the gas flows around the proto-planetesimal, and the particles are dragged with it, they are not returned to the surface.

Particles with a larger stopping time that are ejected against the headwind can reach the region upstream of the proto-planetesimal where the gas velocity points towards the proto-planetesimal. In this region, the particle direction is reversed by the gas flow and it then approaches the planetesimal (trajectory 3). However, because the direction has been reversed, the particle stopping time must consequently still be relatively small. In our simulations, stopping times that lead to a reversal of direction in front of the planetesimal also lead to a complete deflection of the particle and thus no reaccretion. 

For particles with even larger stopping times (trajectory 4), the dust particles leave the collisional cross section of the proto-planetesimal before their motion is reversed. 

In all four cases above, the ejected particles do not collide with the proto-planetesimal, and reaccretion does not occur.

\section{Discussion \& conclusions}
\label{conclusions}
We have examined a number of issues related to the effects of hydrodynamical gas flow around proto-planetesimals, with an interest in the consequences for coagulation efficiency, expanding upon the work of \citetalias{Sek}. By numerically integrating the trajectories of dust particles in the gas flow described by Eq.~(\ref{eq:hydro}), we have quantified how these particles are deflected and their impact velocities affected. We have also studied whether reaccretion of collisional debris remains possible in the presence of a laminar hydrodynamical flow.

We have found that small particles that would impact the proto-planetesimal in the absence of a flow pattern can instead be deflected by the streaming gas and pass over the rim of the proto-planetesimal, avoiding a collision. Even if an impact occurs, the gas flow can decrease the relative velocity between the small particle and the proto-planetesimal, leading to generally less disruptive collisions. These effects are mostly important in a sweep-up scenario, occurring in the regime between small dust aggregates and larger proto-planetesimals, as studied by, \eg, \citet{Xie,Wind}.

Our model is valid for spherical, non-rotating proto-planetesimals with radii $\lesssim 250\, \mathrm{m}$, and within 8~AU of the central star in a MMSN disk, 14~AU in a \citet{Desch} type disk.

The results of our model are summarized as follows:
\begin{itemize}
  \item Sect.~\ref{effects}: The amount of deflection or deceleration experienced by a dust particle as a result of the flow pattern around a proto-planetesimal is purely a function of the parameter $x = R/(v_\infty \tstop)$, where $R$ is the radius of the proto-planetesimal, $\tstop$ is the dust particle stopping time, and $v_\infty$ is the headwind velocity of the disk gas.
  
  \item Sect.~\ref{qualitative}--\ref{sweepupcross}: The flow of disk gas around a proto-planetesimal reduces the effective cross section with which it can sweep up dust particles from the surrounding gas. For a spherical proto-planetesimal, the effective cross section is easily parameterized (Eq.~\ref{eq:sigma}). When $x > x_\mathrm{cutoff} = 0.8$, dust particles no longer impact the proto-planetesimal, and sweep-up is impossible.  This corresponds to a dimensionless stopping time of $1.25$, in rough agreement with \citetalias{Sek}.
  
For example, a proto-planetesimal with a radius $R = 100\, \mathrm{m}$ at 5 AU would, in a MMSN (Desch) type disk with a headwind velocity of $v_\infty = 20\, \mathrm{m\, s^{-1}}$ ($24\, \mathrm{m\, s^{-1}}$), have its effective cross section reduced by half (\ie\ $x \simeq 0.17$ when $D(H) = 4$) for collisions with particles of stopping time $\tstop \sim 30\, \mathrm{s}$ ($25\, \mathrm{s}$), corresponding to a particle size of $0.5$ $\mu$m ($4$ $\mu$m). Particles of stopping time $\tstop \lesssim 6.2\, \mathrm{s}$ ($5.2\, \mathrm{s}$) would not collide with the proto-planetesimal at all ($x \geq x_\mathrm{cutoff}$), corresponding to particle sizes of $\lesssim 0.1$ $\mu$m ($0.9$ $\mu$m).
 
  \item Sect.~\ref{impvel}: The flow pattern reduces the impact velocity of small dust particles onto the proto-planetesimal, and this effect can be parameterized with Eq.~\ref{eq:vimp}. The reduced impact velocity decreases the erosion efficiency, and could result in enhanced sticking.

  \item Sect.~\ref{reaccretefree}: A flow pattern of parallel stream lines in free molecular flow leads to enhanced reaccretion, in agreement with \citetalias{Sek}. However, because the random thermal motion of gas molecules in the disk will dominate over the headwind, parallel streamlines are not physically justifiable for a proto-planetesimal with radius comparable to the mean free path of the gas.

  \item Sect.~\ref{reaccretelaminar}: A laminar flow pattern around a spherical proto-planetesimal does not result in an enhanced reaccretion rate, in agreement with \citetalias{Sek}. In the cases studied here, collisional debris of different stopping times is either sufficiently deflected by the flow to be advected past the proto-planetesimal, or moves beyond the effective cross section before it can be reaccreted.
\end{itemize}

The adopted flow pattern, Eq.\ \ref{eq:hydro}, is strictly valid only for $\ReN \ll 1$, but should provide accurate quantitative results for the sweep-up of dust particles by a spherical proto-planetesimal if $\mathrm{Re} \lesssim 2$ (Sect.\ \ref{sub:realitycheck}).  A relatively simple flow pattern was chosen here to provide a straightforward approximation of the effects of laminar hydrodynamic flow on the accretion of dust particles.  Of course, the dependence of the flow pattern on the Reynolds number influences the results, and while we have explored three analytical approximations (Stokes, Oseen, and Proudman \& Pearson), only full hydrodynamical simulations of the Navier-Stokes equations over the broad parameter space presented here can definitively determine the effects. These simulations will be the subject of future work.

Additional and obvious refinements to our model would encompass surface irregularities, velocity shear of the disk gas, and a non-spherical or rotating proto-planetesimal. Although a non-spherical, or rotating proto-planetesimal, or a shear velocity could produce significant differences in the hydrodynamical flow pattern (\eg\ \citealt{Rotsphere,Ormel2013}), we expect that surface irregularities of planetesimals will introduce only minor differences.

For proto-planetesimals of size $R \sim 100$ m at $5$ AU, our results predict that hydrodynamical deflection and deceleration will considerably reduce the effect of monomeric erosion \citep{Schraep}.  Meanwhile, the detailed consequences of including the reduced sweep-up cross section and impact velocities into a dust growth code are not easily predicted.  The relatively low Reynolds numbers ($\leq 22$) used in this study prevent us from generalising to larger proto-planetesimal radii at low heliocentric distances, and therefore to larger dust particle stopping times.  Even when considering the results of \citetalias{Sek} ($\ReN = 50$ and cut-off point $x \sim 3$), the high Reynolds numbers at low $r$ in the Desch \& MMSN disk models ($> 3000$; Fig.\ \ref{fig:reynoldsnumber}) make it difficult to predict what effect a $\sim\! 100\, \mathrm{m}$ size body might have on the deflection and deceleration of dust particles at $r \sim 1\, \mathrm{AU}$.  This further underscores the need for direct numerical simulation of dust particle sweep-up at high Reynolds numbers.

This study has sampled a broad parameter space and demonstrated that a hydrodynamical gas flow can significantly affect the ability of a proto-planetesimal to sweep-up smaller particles. To further quantify these effects, Eq.~(\ref{eq:sigma}) for the effective sweep-up cross section and Eq.~(\ref{eq:vimp}) for the impact velocities must be implemented into a dust coagulation code which treats sweep-up growth.  In this regard, the appropriate value of $D(H)$ to use in coagulation models is the asymptotic limit of $4$ (\ie\ the dust particles are already well-coupled to the gas flow).
\begin{acknowledgements}
We thank the anonymous referee for a careful and thorough review which improved the manuscript.  JPR is supported by DFG grant DU 414/9-1. FW is funded by the DFG within Forschergruppe 759 “The Formation of Planets: The Critical First Growth Phase”.
\end{acknowledgements}
%
%
\bibliographystyle{aa}
\bibliography{./sellentinetal_v8}
%
%
\end{document}